\documentclass[10pt]{IEEEtran}

\usepackage[all]{xy}
\usepackage{floatflt}
\usepackage{latexsym}
\usepackage{subfigure}
\usepackage{graphicx}
\usepackage{epsfig}
\usepackage{times}
\usepackage{breakurl}
\usepackage[hyphens]{url}
\usepackage{boxedminipage}
\usepackage{ifthen}
\usepackage{caption2}
\usepackage{xspace}
\usepackage{cite}

\newcommand{\capt}{captcha\xspace}
\newcommand{\capts}{captchas\xspace}

\begin{document}

\title{Three-Way Dissection of a Game-CAPTCHA: Automated Attacks, Relay Attacks, and Usability}


\author{
Manar Mohamed$^1$
\and
Niharika Sachdeva$^2$\thanks{Part of the work done while visiting UAB}
\and
Michael Georgescu$^1$
\and
Song Gao$^1$
\and
Nitesh Saxena$^1$\thanks{Contact author: saxena@cis.uab.edu}
\and
Chengcui Zhang$^1$
\and
Ponnurangam Kumaraguru$^2$
\and
Paul C. van Oorschot$^3$
\and
Wei-Bang Chen$^4$\\
$^1$ Computer and Information Sciences, University of Alabama at Birmingham, USA\\
$^2$ Indraprastha Institute of Information Technology, India\\
$^3$ Computer Science, Carleton University, Canada\\
$^4$ Math and Computer Science, Virginia State University, USA
}
\maketitle

\begin{abstract} Existing captcha solutions on the Internet are a major source of user
frustration.
Game {captcha}s are an interesting and, to date, little-studied approach
claiming to make captcha solving a fun activity for the users. One broad form
of such {captcha}s -- called \textit{Dynamic Cognitive Game ({DCG}) {captchas}}
-- challenge the user to perform a game-like cognitive task interacting with a
series of dynamic images. 
%
We pursue a comprehensive analysis of a representative category of {DCG}
{captcha}s. We formalize, design and implement such captchas, and dissect them
across:
(1) fully automated attacks, (2) human-solver relay attacks, and (3) usability.
Our results suggest that the studied DCG captchas exhibit high usability and,
unlike other known captchas, offer some resistance to relay attacks, but they
are also vulnerable to our novel dictionary-based automated attack.  

\end{abstract}

\section{Introduction}
\label{sec:intro}

The abuse of the resources of online services using automated means, such as 
denial-of-service or password
dictionary attacks, is a common security problem.
To prevent such abuse, a primary defense mechanism is {CAPTCHA}
\cite{captcha} (denoted ``captcha''), a tool aimed to distinguish a human user from a
computer based on a task that is easier for the former but
much harder for the latter. 

The most commonly encountered captchas today take the form of a garbled
string of words or characters, but many other variants 
have also been proposed \cite{hidalgo:captchas}. 
Unfortunately, existing captchas suffer from several problems. 
\textit{First}, successful automated attacks have been developed against many existing schemes. 
For example, algorithms have been designed that can
achieve character segmentation with a 90\% success rate \cite{low_cost}. Real
world attacks have also been launched against captchas employed by
Internet giants 
\cite{paypal_attack,microsoft_attack,google_attack}.

\textit{Second}, low-cost attacks have been conceived whereby challenges are relayed to, and solved by, users on
different web sites or paid human-solvers in the crowd \cite{captcha_relay, relay2, relay3}. In fact, it has been shown that \cite{savage-solvers}
such relay attacks are much more viable in practice than automated attacks
due to their simplicity and low economical costs. 

\textit{Third}, the same distortions that are used to hide the underlying content of
a captcha puzzle from computers can also severely degrade human usability
\cite{captcha_usability,bursztein:how-good-are-humans-at-so:2010:zlygf}. More
alarmingly, such usability degradation can be so severe on many occasions that
users get frustrated and give up using the services
that deploy captchas. Consequently, companies lose customers and suffer economic losses \cite{captcha-lose-business}.  

Given these problems, there is an urgent need
to consider alternatives that place the \textit{human user at the center of the
captcha design}.  \textit{Game captchas} offer a promising approach by attempting to make
captcha solving a fun activity for the users. These are challenges that
are built using games that might be enjoyable and easy to play for humans, but hard for computers.


In this paper, we focus on a broad form of game captchas, called
\textit{Dynamic Cognitive Game (DCG)} captchas. This captcha 
challenges a user to perform a \textit{game}-like \textit{cognitive} task
interacting with a series of \textit{dynamic} images. Specifically, we consider a representative DCG captcha category which 
involves objects floating around within the images, and the user's task
is to match the objects with their respective target(s) and drag/drop them
to the target location(s).  
%
A startup called ``are you a human'' \cite{areyouahuman,ayh-deploy} has
recently been offering such \textsc{DCG} captchas.

Besides promising to significantly improve user experience, \textsc{DCG}
captchas are an appealing platform for touch screen enabled mobile devices
(such as smartphones). Traditional captchas are known to be quite difficult on
such devices due to their small displays and key/touch pads, while touch screen
games are much easier and already popular.  Motivated by these unique and
compelling advantages of \textsc{DCG} captchas, we set out to investigate their
security and usability. Specifically, we pursue a comprehensive study of
\textsc{DCG} captchas, analyzing them from three broad yet intersecting
dimensions: (1) \textit{usability}, (2) \textit{fully automated attacks}, and
(3) \textit{human-solver relay attacks}.
%
%
{Our main {contributions}} are as follows:
%
\begin{enumerate}

\item We formalize, design and implement four instances of a representative category of \textsc{DCG} captchas. (\textit{Sections \ref{sec:back} and \ref{sec:design}})

\item We conduct a usability study of these instances, evaluating them 
in terms of time-to-completion, error rates and perceived usability. Our results indicate 
the overall usability to be very good.
(\textit{Section \ref{sec:usability}})

\item We develop a novel, fully automated framework to attack these \textsc{DCG} captcha instances based on
image processing techniques and principles of unsupervised learning. The attack
is computationally efficient and highly accurate, but requires building a dictionary to be effective.
(\textit{Section \ref{sec:comp}})

\item 

We explore different variants of human-solver relay attacks against DCG
captchas.  Specifically, we show that the most simplistic form of relay attack
(in line with traditional captcha relay attack) reduces to a \textit{reaction
time task} for the solver, and conduct a user study to evaluate the performance
of this attack.  In general, our results 
indicate that DCG captchas with mobile answer objects offer some level of
resistance to relay attacks, differentiating them from other captchas. Our
user study may also be of independent interest in other human-centered computing domains.
(\textit{Section \ref{sec:relay}}) 


\end{enumerate}




\section{Background}
\label{sec:back}


We use the term Dynamic Cognitive Game (DCG) captcha to define the broad
captcha schemes that form the focus of our work.
We characterize a \textsc{DCG} captcha as having the following features: (1)
\textit{dynamic} because it involves objects moving around in image frames; (2)
either \textit{cognitive} because it is a form of a puzzle that relates to the
semantics of the images or \textit{image recognition} because it involves
visual recognition; and (3) a \textit{game} because it aims to make captcha
solving task a fun activity for the user.
In this section, we discuss the security model and design choices for DCG
captcha, and present the DCG captcha category and associated instances studied in
this paper.



\subsection{Security Model and Design Choices}
\label{sec:model}
The \textsc{DCG} captcha design objective is the same as that of captcha: a bot (automated computer program) must only be able to solve
captcha challenges with no better than a negligible probability (but a
human should be able to solve with a sufficiently high
probability).\footnote{
%
For example, target thresholds 
might limit bot success rates below
0.6\% \cite{Zhu:2010}, and human user
success rates above 90\% \cite{chellapilla}.} 

A pre-requisite for the security of a DCG captcha implementation (or any
captcha for that matter) is that the responses to the challenge must not be
provided to the client machine in clear text. For example, in a character
recognition captcha, the characters embedded within the images should not be
leaked out to the client.  To avoid such leakage in the context of DCG
captchas, it is important to provide a suitable underlying game platform for
run-time support of the implemented captcha.
Web-based games are commonly developed using Flash and HTML5 in conjunction
with JavaScript.  However, both these platforms operate by downloading the game
code to the client machine and executing it locally. 
Thus, if these game platforms were directly used to implement DCG captchas, the
client machine will know the correct objects and the positions of their
corresponding target region(s), which can be used by the bot
to construct the responses to the server challenges relatively easily. This
will undermine the security of DCG captchas.

The above problem can be addressed by employing two implementation strategies
under different security models. \textit{Model 1} involves
encryption and obfuscation of the game code which will make it difficult for the
attacker (bot) on the client machine to extract the game code and thus the correct
responses. Commercial tools, such as SWF Encrypt \cite{swf-encrypt}, exist
which can be used to achieve this functionality. This approach works under a
security model in which it is assumed that the 
bot does not have the capability
to learn the keys used to decrypt the code and to deobfuscate the code. A similar model where the attacker has
only partial control over the client machine has also been employed in prior
work \cite{weak-captcha-model}.  \textit{Model 2}, on the other
hand, does not restrict the capability of the bot and allows it to
completely control the client machine. It draws from the paradigm
of  \textit{cloud gaming} \cite{cloud-game}, which keeps all the game logic and
code on the server side, and involves the server streaming the game
output to the client machine synchronized with any user input. Products
offered by companies, such as StreamMyGame \cite{streammygame} and Onlive
\cite{onlive}, demonstrate the feasibility of this approach.

The choice of game platform for a DCG captcha implementation depends upon the
desired level of security and performance. The encryption/obfuscation approach
is more efficient but provides a weaker level of security. The game streaming
approach is expected to be secure but may suffer from latencies due to
continuous streaming. It also imposes certain overhead on the captcha
server given that it uses ``thin'' clients.
In practice, we envision a game captcha service to deploy a
hybrid solution whereby the first approach is used on medium/high latency clients and
the second approach is used on low latency clients (assuming latencies can not be faked).

In our model, we assume that the implementation provides continuous feedback to the
user as to whether the objects dragged and dropped to specific target region(s)
correspond to correct answers or not. 
The server also indicates when the game successfully finishes, or times out. 
This
feedback mechanism is essential from the usability perspective otherwise the
users may get confused during the solving process.  The attacker is
free to utilize all of this feedback in attempting to solve the challenges, but
within the time-out. We also assume that it is possible for the server to
preclude brute force attacks, such as when the attacker tries to drag and drop
the regions within the image exhaustively/repeatedly so as to complete the game
successfully. Such a detection is possible 
by simply capping the number of drag/drop attempts per moving object.\footnote{The ``are you a human'' DCG captcha implementation claims to
adopt a sophisticated (proprietary) mechanism, based on mouse events, to
differentiate human game playing activity from an automated activity. 
We did not implement such a human-vs-bot behavioral analysis component because our paper's goal is to examine the underlying captcha scheme only. A behavioral component can be added to other captchas also and represents a topic orthogonal to our work. Besides, it is not clear if behavioral analysis would add security; it may instead degrade usability by increasing false negatives.}




In addition to automated attacks, the security model for DCG captchas (and any
other captcha) must also consider human-solver relay attacks
\cite{captcha_relay,savage-solvers}. In fact, it has been shown that
such relay attacks are much more appealing to the attackers than automated
attacks currently due to their simplicity and low cost \cite{savage-solvers}.  In a relay attack, the
bot forwards the captcha challenges to a human user elsewhere on the Internet
(either a payed solver or an unsuspecting user accessing a web-site
\cite{smuggling}); the user solves the challenges and sends the responses back
to the bot; and the bot simply relays these responses to the server.
Unfortunately, most, if not all, existing captcha solutions are insecure under
such a relay attack model.  For example, character recognition captchas are routinely broken via
such relay attacks \cite{savage-solvers}. For DCG captchas to offer better
security than existing captchas, they should provide some resistance to
such human-solver relay attacks (this is indeed the case as we demonstrate in Section \ref{sec:relay}).  
 
\subsection{Game Instances and Parameters}
\label{sec:game-types}

\begin{figure}[t]
\begin{center}
\centering
\subfigure[Ships Game]{
	\includegraphics[width=.46\columnwidth]{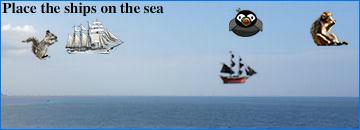}
  	\label{fig:ships}
 }
\centering
\subfigure[Shapes Game]{
	\includegraphics[width=.47\columnwidth]{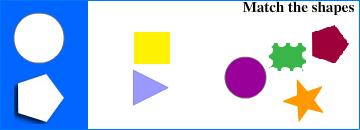}
  	\label{fig:shapes}
 }
\centering
\subfigure[Parking Game]{
	\includegraphics[width=.47\columnwidth]{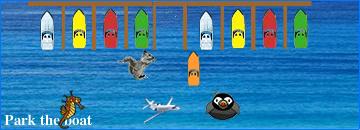}
  	\label{fig:boat}
 }
\centering
\subfigure[Animals Game]{
	\includegraphics[width=.46\columnwidth]{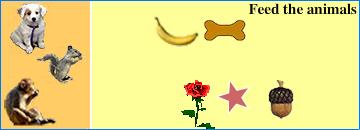}
  	\label{fig:animals}
 }
\caption{Static snapshots of 4 game instances of a representative DCG captcha analyzed in this paper (targets are static; objects are mobile)}
\label{fig:game-cat}
\end{center}
\end{figure}

Many forms of DCG captchas are possible. For example, they may be based on
visual matching or semantic matching of objects, may consist of multiple target
objects or none at all, and may involve static or moving targets.
In this paper, we focus on one representative category, and four associated instances, of DCG captcha with static target(s) (see
Figure \ref{fig:game-cat}).  
Specifically, our studied DCG captcha instances involve:

\begin{enumerate}

\item \textit{single
target object}, such as place the ship in the sea (the Ships game).

\item \textit{two target objects}, such as match the shapes (the Shapes game).

\item \textit{three target objects}, such as feed the animals (the Animals game).

\item \textit{no target objects}, such as park the boat (the Parking game), where
the target area does not consist of any objects. 

\end{enumerate}

The Shapes game is based on visual matching whereas the other games involve
semantic matching. 

For each of these 4 instances, different parameterizations 
affect security and usability. These include: (1) the number of foreground
moving objects, including answer objects and other ``noisy'' objects; and (2)
the speed with which the objects move. The larger the number of
objects and higher the speed, the more difficult and time consuming it might be
for the human user to identify the objects and drag/drop them, which may
degrade usability. However, increasing the number and speed of
objects may also make it harder for a computer program to play the games
successfully, which may improve security.  Thus, for our analysis of the DCG
captcha, we will evaluate the effect of these parameters for
captcha usability and captcha security (against automated as well as relay attacks).  

\section{Design and Implementation}
\label{sec:design}


%
%
%
Due to legal considerations, we did
not resort to directly evaluate an existing DCG captcha implementation (e.g., ``are you a human'' DCG captchas). 
In particular, developing automated
  attacks against these captchas directly violates the company's
  asserted terms and conditions \cite{ayh-terms}.
Instead, we designed and implemented our own
equivalent and generalized versions of DCG captchas from 
scratch, and analyzed these versions.
Developing our own versions also allowed us to freely vary the game parameters, such as the
number and speed of objects, and investigate the DCG captcha security and
usability with respect to these parameters.\footnote{Although our implementation and analyses does not directly involve the ``are
you a human'' captchas, it is generalized enough for our results to be
applicable to these captchas also (i.e., the ones that fall under the
categories evaluated in our work).}  

We created four instances of games as specified
in Section \ref{sec:game-types} using Adobe Flash. 
%
%
This follows
\textit{Model 1} (Section \ref{sec:model}) whereby the games reside locally
on the client machine. However, since our purpose
was to analyze the implemented captcha security against automated (image
processing) attacks and relay attacks (and not for leakage of information from
the client code itself), 
we did not encrypt or obfuscate our captcha
  game implementation. All of our security analysis is also generally applicable to \textit{Model 2}.


The game image/frame size is 360 x 130 pixels, which can easily fit onto a web page
such as in a web form. 
Each game starts by placing the objects in certain pre-specified locations on
the image. Then, each object picks a random direction in which it
will move. A total of 8 directions were used, namely, N, S,
E, W, NE, NW, SE and
SW.  If the chosen direction is one of E, W, S, or N, the object
will move (across X or Y axis) by 1 pixel per frame in that direction.
Otherwise, 
the
object will move $\sqrt{2} = 1.414$ pixels per frame along the hypotenuse,
corresponding to 1 pixel across both X and Y axes.  This means that on an
average the object moves $1.207$ [$=(1*4+1.414*4)/8$] pixels per frame.  
The object continues in the 
  current direction until colliding with another object or the game border, 
  whereupon it moves in a new random direction.

The game starts when the user presses a ``Start'' button on the screen center.
Each game briefly explains to users their task, e.g., ``Place the ships on the
sea." The game ends when the user clicks/drags all the correct objects onto
their corresponding target(s), in which case a ``Game Complete'' message is
provided. 
To successfully match an object with its target, the user clicks inside the
bounding box across the shape of the object, drags the object and drops it by releasing 
it inside the bounding box across the respective target.
The game must be successfully
  completed within a fixed time (we allow 60s); the user gets  
  feedback on the correctness of every drag-drop, by a star on 
  success and a cross on failure (Figure \ref{fig:feedback}, Appendix \ref{sec:app-fig}).


For each of the 4 games, we set 5 parameterizations, choosing object speed
(low, medium, high) as (10, 20, 40) frames per second (FPS), and number of
moving objects as (4, 5, 6).  (These frame rates translate into average
object speeds of 12.07, 24.14 and 48.28 pixels/second, resp., given the
objects move $1.207$ pixels/frame.) For each game, we used 5
combinations of speed and number of objects: (10 FPS, 4 objects); (20 FPS, 4 objects); (20 FPS, 5 objects); (20 FPS, 6 objects); and (40 FPS, 4 objects). This resulted in a total of 20 games in our corpus.


\section{Usability}
\label{sec:usability}

In this section, we report on a usability study of our representative DCG captcha category.

\subsection{Study Design, Goals, and Process}

Our study involved \textbf{40 participants} who
  were primarily students from various backgrounds.
(For demographics, see Table
\ref{tab:demographics-interviews}). The study was
\textit{web-based} and comprised of three phases. The \textit{pre-study phase}
involved registering and briefly explaining the participants about the
protocols of the study.  In particular, the participants were shown ``consent
information,'' which they had to agree before proceeding with the study.  This
was followed by collecting participant demographics and then the participants
playing the different DCG captcha games.  This \textit{actual study phase}
attempted to mimic a realistic captcha scenario which typically involves
filling out a form followed by solving captcha. To avoid explicit priming,
however, we did not mention that the study is about captcha or security, but
rather indicated that it is about assessing the usability of a web interface.
In the \textit{post-study phase},
  participants answered questions about their experience with the tested DCG captchas.
This comprised the standard SUS (Simple Usability Scale)
questions \cite{sus}, a standard 10-item 5-point Likert
scale (`1' represents ``Strong disagreement'' and `5' represents ``Strong agreement''). SUS polls satisfaction with respect to computer systems \cite{sus-chi}, in order to
assess their usability.  Additionally, we asked
several other questions related to the games' usability.

\begin{table}[h]
{\footnotesize{
\caption{Usability Study Participant Demographics}
\begin{center}
\small
\begin{tabular}{|p{5cm}| r |}
\hline
\textbf{}& \textbf{N=40} \\
\hline
\hline
\textbf{Gender} & (\%) \\
\hline
 Male & 50\\
 \hline
Female& 50 \\
\hline
 \hline
\textbf{Age} & (\%) \\
\hline

18 - 24 & 80 \\
 \hline
 25 - 35 & 20 \\
 \hline
  \hline
\textbf{Education} & (\%) \\
\hline 
Highschool & 45 \\
 \hline
Bachelors  & 27.5\\
 \hline
Masters & 22.5\\
\hline
Ph. D. & 5 \\
 \hline
  \hline
\textbf{Profession / field of study} & (\%) \\
  \hline
Computer Science & 60 \\
  \hline
Engineering & 5 \\
 \hline
Science, Pharmaceuticals & 10\\
 \hline
 Law & 2.5\\
 \hline
Journalism  & 2.5 \\
  \hline
Finance  & 2.5 \\
  \hline
Business  & 5 \\
  \hline
 Others  &12.5 \\
  \hline
\end{tabular}
\normalsize
\end{center}
\label{tab:demographics-interviews}
}}
\end{table}



In the actual study phase, each participant
  played 20 instances as discussed in Section \ref{sec:design}, aimed at understanding how
  different parameterizations impact users' solving capabilities. The
  order of games presented to different participants involved a
  standard 20x20 Latin Square design to counter-balance learning effects.
  Via our study, our goal was to assess the following aspects of the DCG captchas:

\begin {enumerate}

\item \textit{Efficiency}: time taken to complete each game.
\item \textit{Robustness}: likelihood of not completing the game, and of incorrect drag and drop attempts.
\item \textit{Effect of Game Parameters}: the effect of the object speed and number on completion time and error rates.
\item \textit{User Experience}: participants' SUS ratings and qualitative feedback about their experience with the games.

\end {enumerate}

For each tested game, completion times, and errors
were automatically logged by our web-interface software.

\subsection{Study Results}

We now provide the results from our usability study, including time to
completion and error rates, as well as perceived qualitative aspects of the
methods based on user ratings. 

\vspace{1mm}
\noindent {\bf Completion Time}: Table \ref{tab:captchaeff} shows
the completion time per game type. Clearly, all games
turned out to be quite fast, lasting for less than 10s on an average.
Users took
longest to solve the Animals game with an average time of 9.10s,
whereas the other games took almost half of this time.  This might have been
due to increased semantic load on the users in the Animals game to identify
three target objects and then match them with the corresponding answer objects.
Moreover, we noticed a decrease in the solving time (equal to 3.84s) when
the target objects were decreased to 2 (i.e., in the Shapes game), and this
time was comparable to games which had 1 target object in the challenge (Ships and Parking). A one-way repeated-measures ANOVA
test showed significant difference (at 95\% confidence) in the mean timings of all 4 types of games
($p<0.0001$, $F=79.98$). Aalyzing further using pairwise paired
t-tests with Bonferroni correction, we found significant difference between the mean times of following pairs:
Animals and Parking ($p < 0.001$),
Ships and Shapes ($p<0.0005$),
Animals and Ships ($p<0.001$), 
Animals and Shapes ($p<0.001$), 
and Parking and Shapes ($p=0.0024$).

{\footnotesize{
\begin{table}[ht]
  \begin{center} 
  \caption{Error rates per click and completion time per game type}
    \begin{tabular}{|p{1.4cm}|p{2.8cm}|p{2.8cm}|}
   \hline

{\bf Game Type} &  {\bf Completion Time (s)} \textit{mean (std dev)}& {\bf Error Rate Per Click}   \textit{mean}  \\
 \hline
 \hline
    \raggedright{Ships} &4.51 (1.00) & 0.04 \\ 
 \hline
  \raggedright{Animals} &9.10 (0.96)&0.05 \\ 
\hline
  \raggedright{Parking} &4.37 (0.90) &0.09 \\
  \hline
  \raggedright{Shapes} &5.26 (0.59)  &0.03 \\ 
\hline 
    \end{tabular}
  \label{tab:captchaeff}
\end{center} 
\end{table}%
}}

\noindent {\bf Error Rates}:  An important result is that all the
tested games yielded 100\% accuracy (\textit{overall error rate of 0\%}).  In other
words, none of the participant failed to complete any of the games within the
time out. This suggests our DCG captchas instances are quite robust to human
errors.

Next, we calculated the likelihood of incorrect drag and drop
attempts (\textit{error rate per click}). For example,  in the Animals game, an incorrect attempt would be to
feed the monkey with a flower instead of a banana.  We define the error rate per click as the number of incorrect objects (from the pool of all foreground objects) dragged to the target area \textit{divided by} the total number of objects dragged and
dropped. The results are depicted in Table \ref{tab:captchaeff}.  
We observe that the Shape game yields the smallest average per click error
rate of 3\%.
This suggests that the visual matching task (as in the Shapes game) is less
error prone compared to the semantic matching task (as in the other games). The
game challenge which seemed most difficult for participants was
the Parking game (average per click error rate 9\%). Since objects in this game are
relatively small, participants may have had some difficulty to
identify them.



\vspace{1mm}
\noindent {\bf Effect of Object Speed and Number}: Table
\ref{tab:captchaspeeduser} shows the performance of the game captchas in terms
of per click error rates and completion time as per different object speeds.  
We can see that 
the maximum number of per click errors were committed at 10 FPS speed.
Looking at the average timings, we find that it took longest to complete
the games when the objects move at the fastest speed of 40 FPS, while 20 FPS
yielded fastest completion time followed by 10 FPS.  ANOVA test revealed
statistical difference among the mean completion time corresponding to three speeds
($p=0.0045$, $F=5.65$). Further analyzing using the t-test with Bonferroni correction, we found statistical
difference between the mean timing corresponding to the following pair of speeds only: 10 FPS and 20 FPS ($p=0.0001$).

{\footnotesize{
\begin{table}[ht]
  \begin{center} 
  \caption{Error rates per click and completion time per object speeds}
    \begin{tabular}{|p{1.7cm}|p{2.8cm}|p{2.8cm}|}
   \hline

{\bf Object Speed} &  {\bf Completion Time (s)} \textit{mean (std dev)}& {\bf Error Rate Per Click}   \textit{mean}  \\

 \hline
 \hline
    \raggedright{10 FPS } &5.74 (2.11) &0.06\\ 
 \hline
  \raggedright{20 FPS} &4.90 (2.22) &0.05 \\ 
\hline
  \raggedright{40 FPS} &6.53 (2.87) &0.04 \\ 
  \hline
    \end{tabular}
  \label{tab:captchaspeeduser}
\end{center} 
\end{table}%
}}

Another aspect of the
usability analysis included testing the effect of increase in the number of
objects (including noisy answer objects) on the overall game performance.
Table \ref{tab:captchaobjectuser} summarizes the per click error rates and completion
time against different number of objects. Here, we can see a clear pattern of
increase, albeit very minor, in average completion time and average rate with
increase in the number of objects. This is intuitive because increasing the
number of objects increases the cognitive load on the users which may slow down
the game play and introduce chances of errors.  ANOVA test did not indicate
this difference to be significant, however.

{\footnotesize{
\begin{table}[ht]
  \begin{center} 
  \caption{Error rates per click \& completion time per \# of objects}
    \begin{tabular}{|p{1.7cm}|p{2.8cm}|p{2.8cm}|}
   \hline
{\bf \# of Objects} &  {\bf Completion Time (s)} \textit{mean (std dev)}& {\bf Error Rate Per Click}   \textit{mean}  \\

 \hline
 \hline
    \raggedright{ 6 } &6.58 (1.69) &0.06 \\ 
 \hline
  \raggedright{5 } &5.30 (2.28) &0.05 \\ 
\hline
  \raggedright{4 } &4.90 (2.22) &0.04 \\ 
  \hline 
    \end{tabular}
  \label{tab:captchaobjectuser}
\end{center} 
\end{table}%
}}

\noindent {\bf User Experience:}
Now, we analyze the data collected from the participants during the post-study
phase. The average SUS score came out to be 73.88 (standard
deviation = {6.94}). Considering that the average SUS scores for user-friendly industrial software tends to
hover in the 60--70 range \cite{susmean}, the usability of our DCG game captcha instances can be
rated as high.

In addition to SUS, we asked the participants a few 5-point Likert scale
questions about the usability of the games (`1' means ``Strong Disagreement''). Specifically, we asked if the games
were ``visually attractive'' and ``pleasurable,'' and whether they would like
to use them in ``practice.'' Table~\ref{tab:usersus}, shows the corresponding average
Likert scores. 
We found that 47\% percent participants felt that the games were visually
attractive and 45\% said that it was pleasurable to play the games. These
numbers indicate the promising usability exhibited by the games.
We further inquired users if they noticed change in speed or number of objects
in the games. 27.5\% noticed no change (increase and/or decrease) in speed of
objects, whereas only 22.5\% noticed no change in number of objects (see
Table \ref{tab:userother}). Thus, the change in the
number of objects and speed (within the limits we tested) was 
noticeable by a large fraction of participants.

{\footnotesize{
\begin{table}[ht]
  \begin{center}
        \vspace{-5mm}
  \caption{User feedback on game attributes}
    \begin{tabular}{|p{4cm}|p{2cm}|}
   \hline
{\bf Attribute} &  \bf Likert Score\\
 &  \textit{mean (std dev)}\\
 \hline
 \hline
    \raggedright{Visually Attractive} & 3.18 (0.94)\\
 \hline
  \raggedright{Pleasurable } & 3.33 (0.96)\\
\hline
    \end{tabular}
  \label{tab:usersus}%
  \end{center}
\end{table}%
}}

{\footnotesize{
\begin{table}[ht]
  \begin{center}
  \caption{\% of users noticing change in speed and number of objects}
    \begin{tabular}{|p{4cm}|p{2cm}|}
   \hline
  \textbf{Object Speed} & (\%) \\
 \hline
 \hline
    \raggedright{Moved faster} &30 \\
 \hline
  \raggedright{Moved slower} & 5 \\
\hline
  \raggedright{No change} & 27.5\\
\hline

  \raggedright{Both slower and faster} & 37.5\\
\hline
 \textbf{Number of objects} & (\%) \\
 \hline
 \hline
    \raggedright{Increased} & 47.5\\
 \hline
  \raggedright{Decreased } & 2.5\\
\hline
  \raggedright{No change} & 22.5\\
\hline
  \raggedright{Both increase and decrease} & 27.5\\
\hline
    \end{tabular}
  \label{tab:userother}%
  \end{center}
\end{table}%
}}

\smallskip \noindent {\bf \underline{Summary of Usability Analysis}:} 
Our results suggest that the DCG captcha representatives tested in this work
offer very good usability, resulting in short completion times (less than 10s),
very low error rates (0\% per game completion, and less than 10\% per drag and
drop attempt),\footnote{When contrasted with many traditional captchas
\cite{bursztein:how-good-are-humans-at-so:2010:zlygf}, these timings are
comparable but the accuracies are significantly better.} and good user ratings.
We found that increasing the object speed and number is likely to degrade the
game performance, but up to 6 objects and up to 40 FPS speed yield a good level
of usability. 
Although our study was conducted with a relatively young participant pool,
given the simplcity of the games (involving easy matching and clicking tasks), we anticipate the game performance to be
generally in line with these results.\footnote{For example, ``are you a human'' FAQs also report similar timings for their games, averaging 10-12s \cite{areyouahuman}.}


\section{Automated Attacks}
\label{sec:comp}

Having validated, via our usability study, that it is quite easy for the human users to play our DCG captcha instances, we next 
proceeded to determine how difficult these games might be for the computer programs. In this section, we present and
evaluate the performance of a fully automated framework that can solve DCG
captcha challenges based on image processing techniques and principles of
unsupervised learning.
We start by considering random guessing attacks and then demonstrate that
our framework performs orders of magnitude better than the former. 

\subsection{Random Guessing Attack}
\label{sec:brute-force}

An attacker given a DCG captcha challenge can always attempt to perform a
random guessing attack. Let us assume that the attacker knows which game he is
being challenged with as well as the location of the target area (e.g., the
blue region containing the target circle and pentagon in the Shapes game) and
the moving object area (e.g., the white region in the Shapes game within which
the objects move). Although determining the latter in a fully automated fashion
is a non-trivial problem (see our attack framework below),
an attacker can obtain this knowledge with the help of a human solver.

However, the attacker (bot) still requires knowledge of: (1) the foreground 
objects (i.e., all the objects in the moving object area) and (2) the target
objects (i.e., the objects contained within the target area). A randomized
strategy that the attacker could adopt is to pick a random location on the
moving object area and drag/drop it to a random location on the target area.
More precisely, the attacker can divide the moving object area and the target
area into grids of reasonable sizes so as to cover the sizes of foreground moving 
objects and target objects.  For example, the moving object area can be divided
into a 10 pixel x 10 pixel grid and target region can be divided into
a 3 pixel x 3 pixel grid (given that the target area size is roughly 3 times the object area
size).  If there are a total of $r$ target objects, the total number of
possibilities in which the cells (possibly containing the answer objects) on
the object area can be dragged and dropped to the cells on the target area are
given by $t = C(100, r)*P(9, r)$. This is equivalent to choosing $r$ cells in
the object area out of a total of $100$ cells, and then rearranging them on to
$9$ cells in the target area. Thus, the probability of attacker success in solving the challenge in a single attempt is $1/t$. For the
DCG captcha instances targeted in this paper, $r$
is 3, 2 and 1, resulting in the respective success probabilities of
$0.00000123\%$, $0.000281\%$ and $0.1\%$. Each attempt corresponds to $r$
drag-and-drop events.  Even if the attacker is allowed a limited (3-4) number
of attempts to solve the captcha, these probabilities are still much lower than
the target probabilities for a real-world captcha system security (e.g., 0.6\%
as suggested by Zhu et al.\ \cite{Zhu:2010}). 

While this analysis suggests
that such DCG captchas are not vulnerable to naive guessing attacks,
the next step is to subject them to more sophisticated, fully
automated attacks, as we pursue below.

\subsection{Our Automated Attack and Results} 

Our attack framework
involves the following phases: 

\begin{enumerate}
%
\item 
Learning the background image of the challenge and identifying the foreground 
moving objects. A background is the canvas on which the foreground objects are
rendered. The foreground objects, for example, in the Ships game, as shown in
Figure \ref{fig:ships}, are bird, ship, monkey, and squirrel.
%
\item Identifying the target area and the target area center(s). For example,
the sea in the Ships game, and the animals in the Animals game. 
%
\item Identifying and learning the correct answer objects. For example, the ships in the
Ships game.

\item Building a dictionary of answer objects and corresponding targets, the background image,
the target area and their visual features, and later using this knowledge base to attack
the new challenges.
%
\item Continuously learning from new challenges containing previously unseen objects. 

\end{enumerate}










Next, we elaborate upon our design and matlab-based implementation per each
attack phase as well as our experimental results.  We note that, on a web forum
\cite{ayh-attack}, the author claims to have developed an attack against the
``are you a human'' captcha. However, \textit{unlike our generalized framework}, this
method is perfected for \textit{only one simple game} that has one single target area
and a fixed set of answer objects. It is not known whether or how easily this
method can be adapted to handle different games, games with multiple instances
that carry different sets of answer objects, and those with multiple target
areas. Since only one game is cracked, one needs to keep refreshing the game
page, if allowed, until that specific game appears.  Since no technical details are provided
in \cite{ayh-attack}, we can only doubt if any background learning or object
extraction is implemented by observing the short time it takes to finish the
attack.
 
\smallskip
\noindent {\bf (1) Background \& Foreground Object Extraction:} 
To extract the moving objects in the challenge, we developed a
\textit{background \& foreground extraction technique}. By subtracting the background image
from a video frame, the foreground moving objects become readily extractable.
In our approach, 40 frames captured at a fixed time interval (0.2s) are used
to learn the background image. For each pixel on the game scene, its 40
observed color values are collected and a histogram is built to select the most
frequent color value (dominant color) as the background color for that pixel.
This is based on the assumption that the background image is static and the
foreground objects are constantly moving, such that the true background color
almost always appears as the most frequent (or consistent) color observed for a
pixel. This approach is computationally efficient and sufficiently accurate for
our purpose. To further reduce the computational cost, a 6-bit
color code
rather than a 24-bit or
3-byte representation of a color value, is used to code the video frame, the
learned background image, and the learned foreground objects. 

Each learned background image is saved in the database. After removing the
extracted background from 5-8 equally distant frames from the 40 initial
frames, the objects in each of the selected frame are extracted. The objects
below a certain size threshold were discarded as noise. The frame with the
maximum number of objects was then selected to extract
various objects. Using multiple frames for object extraction also helped us
discard the frames in which the objects overlapped each other and were hence
detected as a single object instead of distinct individual objects.

\begin{figure}[t!]
\begin{center}
\vspace{-.5in}
\includegraphics[width=\columnwidth]{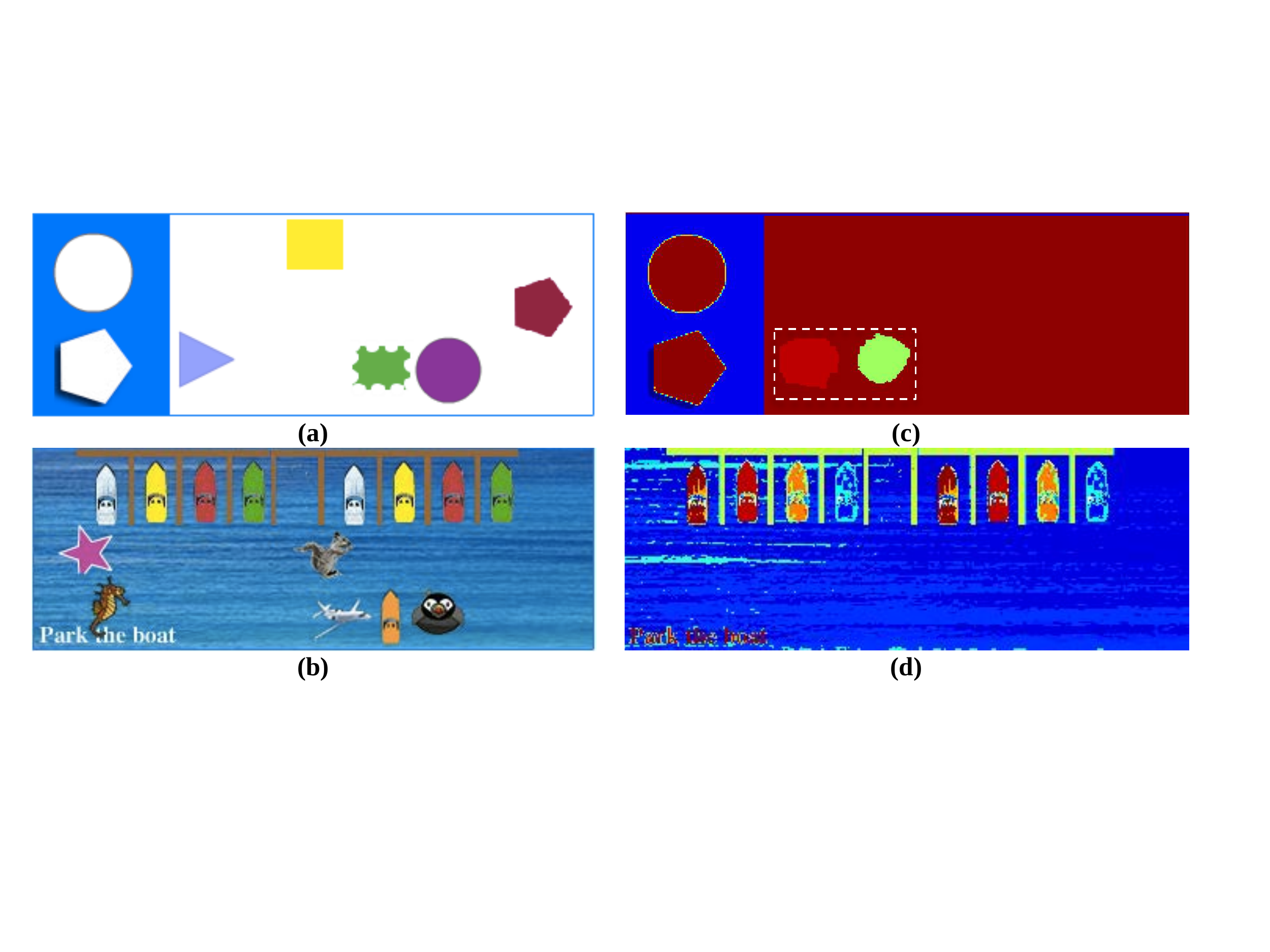} 
\vspace{-.8in}
\caption{Detected backgrounds. (a)(b) The original game scenes of the Shapes and Parking challenges; (c)(d) Detected background images in color codes}
\vspace{-2mm}
\label{fig:background-fig2}
\end{center}
\end{figure}

Figure \ref{fig:background-fig2} shows two detected background images that are represented in color
codes of the Shapes and the Parking games. One drawback of
this method is that if the moving speed of the foreground objects is too slow,
especially when some foreground objects hover over a small area, the dominant
color values of most pixels in that area will be contributed by the foreground
objects instead of by the background. A shadow of foreground objects may appear
as a pseudo patch in the background image as shown in Figure \ref{fig:background-fig2}(c), indicated
by a dashed rectangle. According to our experimental results, the likelihood of
observing such pseudo patches is sufficiently low ($<$ 7\%). Even if there is any,
one possible solution is to use more frames for initial background learning.
However, the time cost is directly proportional to the amount of frames
collected. On the other hand, pseudo patches may not pose a big issue. Even
though the existence of pseudo patches may result in over-segmented foreground
objects when they overlap each other, a partially detected object can still be
used to extract visual features and later to locate an object that matches the
visual features at the time of attacking. Yet there is another possible
solution that uses multiple running instances of the same game to learn the
background. The idea is to take full advantages of the different initial
configurations of a game and utilize the variety in the set to construct a more
reliable background image. The overall learning time will be increased but the
individual learning time for each instance remains the same, which avoids the
failure case in which the game stops before the background is fully
learned.

As the final step as part of this phase, the visual features, coded as color code
histograms (a visual feature commonly used to describe the color distribution
in an image), of the foreground objects and the background image, are
stored in the database, together with some other meta-data such as the object
size and dimensions.

\smallskip
\noindent {\bf (2) Target Area Detection:} 
Identifying the target area requires analysis of the
background extracted in the previous phase. For this purpose, we implemented two
alternative approaches, namely the \textit{Minimum Bounding Rectangle (MBR)} \cite{mbr} method and the
\textit{Edge-based method}, and compared and contrasted them with regard to detection
accuracy and time efficiency.

\begin{figure}[h]
\begin{center}
\vspace{-.35in}
\includegraphics[width=\columnwidth]{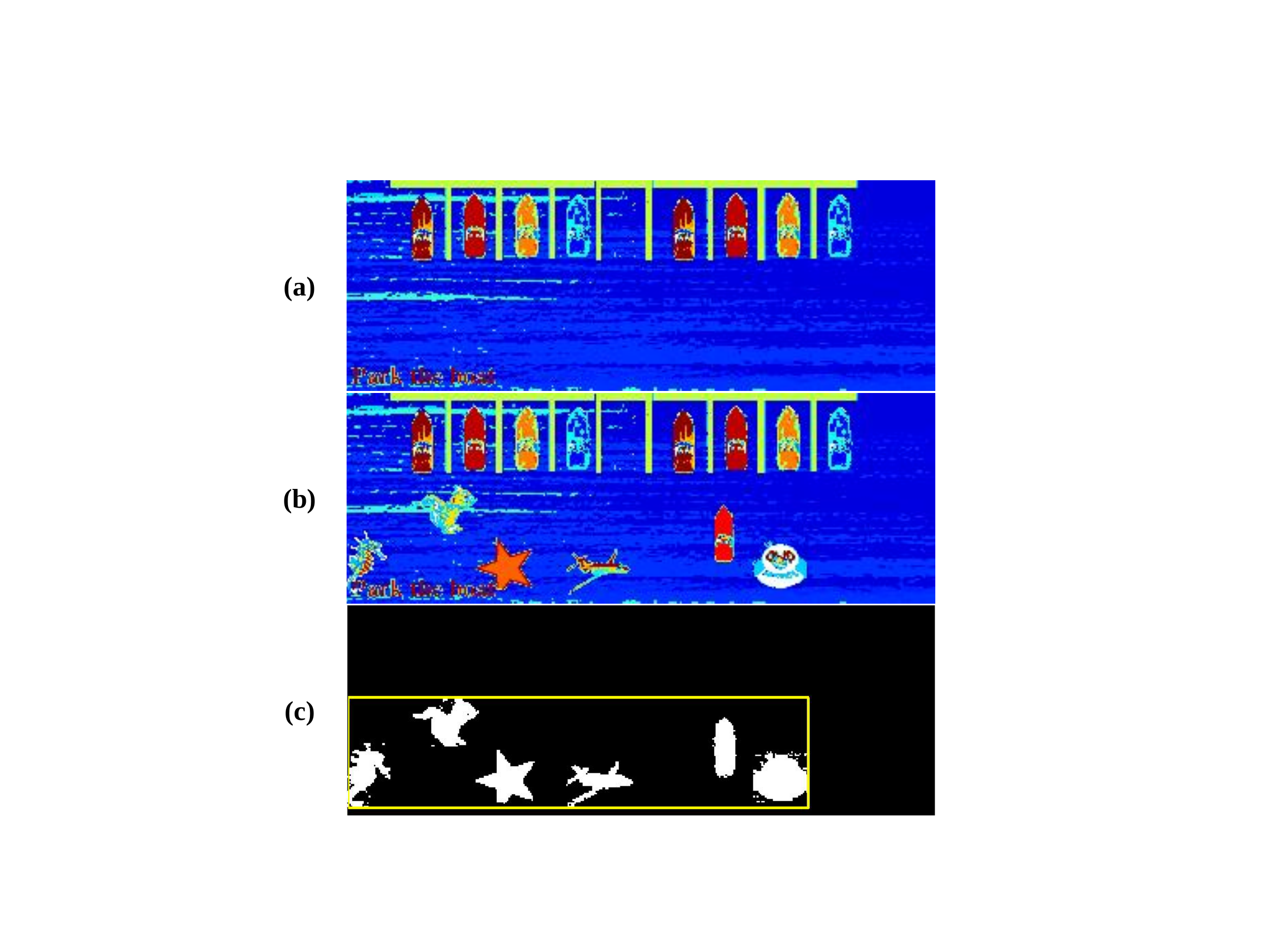} 
\caption{Target Detection. (a) The detected background for the Parking challenge; (b) One sample frame represented in color code; (c) Detected foreground objects from (b) and their MBR.}
\label{fig:target-detection}
\end{center}
\end{figure}

The MBR-based method is based on the observation that the
activity/moving area of foreground objects has no or very little overlap with
the target area. Therefore, by detecting and removing the foreground moving
area from the background image, a reasonable estimate of the target area can be
obtained. As the first step of this approach, the selected 5-8 frames and their
foreground object masks from the previous phase are used to identify the
foreground moving area mask. More specifically, the foreground mask is
generated by identifying those pixels that have a different color code value
than that of the corresponding pixels in the background image. Then, a Minimum
Bounding Rectangle (MBR) is generated that bounds the area where the foreground
objects are detected in the current frame (Figure \ref{fig:target-detection}). The final
estimate of the foreground moving area, denoted as $MBR_{final}$, is the
superimposition of all the MBRs extracted from the sample frames, also
represented as a minimum bounding rectangle (see Figure \ref{fig:target-detection}(c)).

\begin{figure}[h]
\begin{center}
{\includegraphics[width=.3\textwidth]{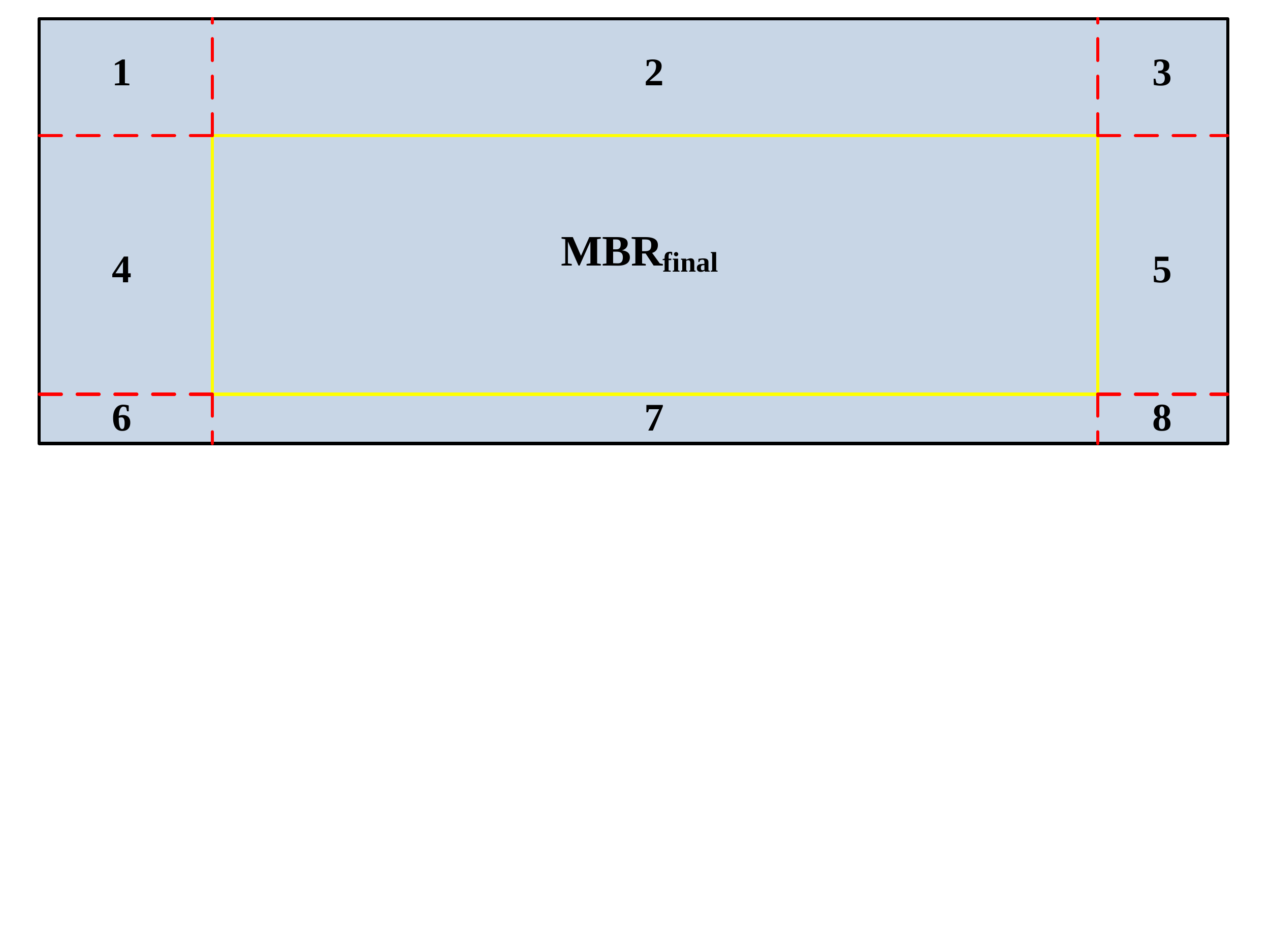}} 
\vspace{-.8in}
\caption{Eight sub-areas generated according to moving area of foreground objects} \label{fig:mbr}
\end{center}
\end{figure}

After the removal of the entire area bounded by $MBR_{final}$ from the background
image, the remaining background is divided into eight sub-areas as shown in
Figure \ref{fig:mbr}. The sub-area with the largest area (e.g., sub-area \#2 in
Figure \ref{fig:mbr}) is identified as the target area, and its centroid is the target
center. It is worth noting that the computational cost of this method is very
low ($O(MN)$, where $N$ is the number of pixels in a game scene, $M$ is the number
of sample frames, and $M\ll N$) since the foreground object masks are readily
available as part of the output from the previous phase. In other words, the
most time consuming part, which is the extraction of foreground 
objects ($O(MN^2)$) from sample frames, has been covered in the previous phase.\footnote{We also implemented an alternative design, called the \textit{exclusion
method} (see Appendix \ref{sec:exclusion}), which detects the target area by simply removing foreground object
pixels accumulated from all the sample frames. However, while this method
is slightly faster than the MBR-based
method, it is less robust.}

The Edge-based method employs a different design principle than MBR-based method.
It is
based on the hypothesis that there are strong edges in the target area
because of the likely presence of objects in the target area, such as the dog
and the squirrel in the Animals game. The steps involved in the edge-based
method are listed below:

\begin{enumerate}

\item Collect a sequence of frames and learn the background image as in the MBR-based method.
\item Detect edge pixels on the background image. Group connected edge pixels into edge segments.
\item Remove trivial edge segments that have too few pixels by a user-input threshold.
\item The mean of all the centroids of remaining segments is used as the target area center.

\end{enumerate}

\begin{figure}[t]
\begin{center}
\vspace{-.3in}
\includegraphics[width=.5\textwidth]{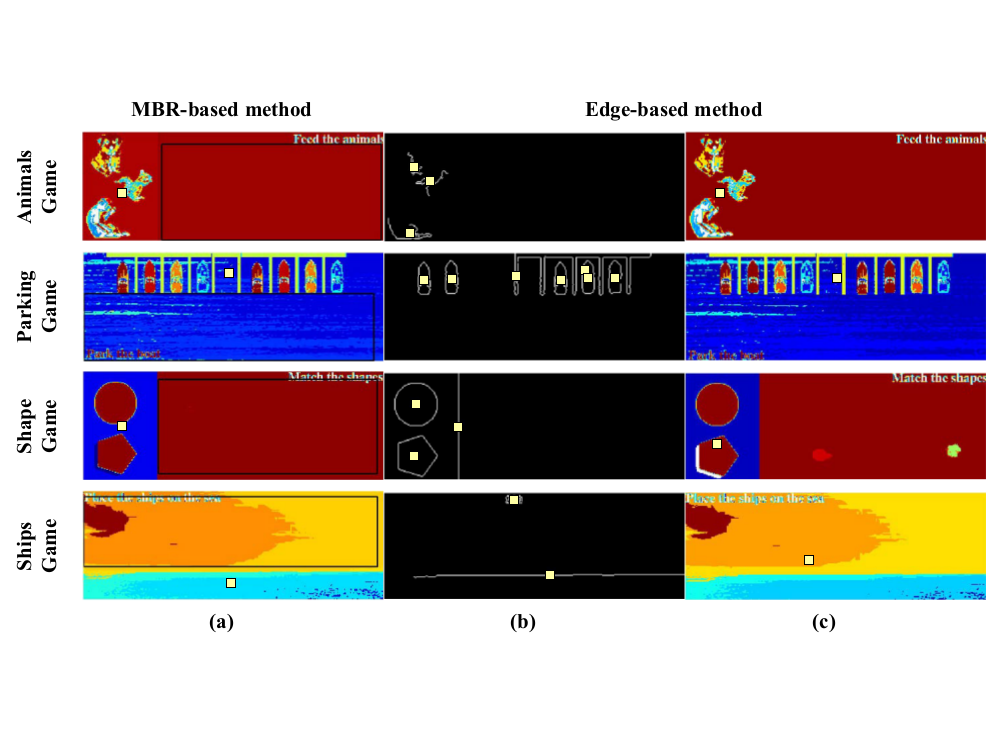} 
\vspace{-.4in}
\caption{Comparison of the target area center detection results between the MBR-based and the edge-based methods. (a) Results from MBR-based method (solid square dot represents the target area center and black rectangle represents the object moving area); (b) centroids of non-trivial edges from the edge-based method; and (c) final target area centers from the edge-based method.}
\vspace{-2mm}
\label{fig:mbs-and-edge}
\end{center}
\end{figure}

The comparison results of the MBR-based and Edge-based methods are shown in Figure
\ref{fig:mbs-and-edge}. The solid square dot in each game scene in Figure
\ref{fig:mbs-and-edge}(a) is the MBR-detected target area center for that
challenge. Also displayed in Figure \ref{fig:mbs-and-edge}(a) are the detected foreground object
moving areas, namely $MBR_{final}$, displayed as a black rectangle in each game
scene. According to our experimental results, MBR-based method was able to
detect the correct target area center in all the challenges. In contrast, for
the edge-based method, it is difficult to find a global threshold that works
for all the challenges. Rather, we need to adjust the threshold for a specific
game in order to achieve ``reasonably good'' results, and this method is also
sensitive to the existence of texts in the background. Figure
\ref{fig:mbs-and-edge}(b) shows the ``optimal'' edge detection result for each
challenge with a manually tuned threshold which is different for each
challenge. As shown in Figure \ref{fig:mbs-and-edge}(c), some target area
centers are incorrectly detected because some edge segments belong to the texts
that are part of the background but not of the target area. This means that the
accuracy of the edge-based method could be significantly undermined by the
presence of strong edges in the background that are not part of the target area
(e.g., presence of texts) and the absence of objects in the target area (e.g.,
the absence of objects in the target area of the Ships Game). As for efficiency, the MBR-based method has a time complexity 
of $O(MN)$
where $M$ is a constant in the range of 5-8, 
while the time complexity of the
edge-based method is 
$O(NL+N^2)$ where $L$ is a constant in the range of 3-8
estimated based on the typical time complexity of a non-combining edge
detection method 
\cite{ref-3}. Overall, this shows that the MBR method
outperforms the Edge method on several aspects.

\vspace{1mm}
\noindent {\bf (3) Answer Object \& Target Location Detection}: Once the target
area is identified, the next step is to identify the correct answer objects and
their respective matching sub-target areas. Since a game can not have too many
sub-target areas (otherwise, usability will be compromised), we divide the
entire probable target area into 9 equal-sized blocks, each represented by its
area centroid, drag each foreground object to each of the 9 centroids, and stop
and record the knowledge learned whenever there is a ``match.'' A match occurs
when an answer object is dragged to its corresponding sub-target area (e.g., a
``bone'' dragged onto a ``dog''). This is detected by monitoring the change of
the area summation of all the foreground objects, since an answer object, once
dragged to its correct target location, will stay in the target area and
therefore result in a reduction of the foreground area. In our experiments,
this method has proven 100\% effective when applied to all four games. As for
efficiency, while the worst case upper bound is $O(N)$, where $N$ is the total
number of foreground objects, in practice, much less number of drags are
required. Our experimental results show that, with 5 foreground objects for
each game (the maximum setting) and 15 training runs for each game, the average
number of drags needed for a game is 9, i.e., less than 2 drags per each object
on average.  In case the server imposes a strict limit on drag/drop attempts, this process can be repeated over mutiple runs.

\vspace{1mm}
\noindent {\bf (4) Knowledge Database Building and Attacking:} The background, target area, and learned answer objects as
well as their corresponding sub-target areas together constitute the knowledge
database for a game. After learning about sufficient number of games,
whenever a new game challenge is presented, the knowledge base is checked for
the challenge. The target area of the currently presented challenge is matched
with the target areas present in the database to identify the challenge. If a
match is found, the extraction of objects from the foreground follows. The
visual features such as the color code histogram of the currently extracted
objects are matched with that of the answer objects in the database for that
challenge.  The extracted objects identified as correct answer objects are then
dragged to their corresponding sub-target areas. To measure the performance of
our approach, we ran this attacking module 100 times for each game instance,
and the average successful attacking time is 6.9s with the number of foreground
objects ranging from 4 to 6. The maximum successful attacking time is 9.3s,
observed for an instance of the Animal game with 6 foreground objects. These
timings are in line with those exhibited by honest users in our
usability study, which will make it impossible for the captcha server to
time-out our attack.  

\vspace{1mm}
\noindent {\bf (5) Continuous Learning:}
During attacking, if a challenge matches a game in our database but contains
previously unseen answer object(s) (e.g., a new ship object in a Ships game
instance), the attack will not terminate successfully. Whenever such a
situation arises, an answer object learning module that is similar to the
aforementioned module is activated, but differs from the latter in that it only needs
to drag a potential answer object to each of the previously learned sub-target
areas that have matching answer objects in the database. The newly learned
answer objects and their corresponding sub-target area centroids are then
added to the knowledge base for that game.

\vspace{-2mm}
\subsection{Discussion and Summary}
\label{sec:attack-disc}

Though background learning is the most time-consuming process in our framework
(it took on an average 30.9s when we ran this module 15 times per game
instance), there are two benefits that it provides.  First, the learned
background can be used to quickly extract foreground moving objects. 
Second, the learned background can be used to locate the target area where
foreground answer objects need to be dragged to.
%
%
%
%
%
%
Our background learning
method has a fixed learning time, but further improvements on both the
efficiency and accuracy can be devised. For example, the learning can be
terminated whenever sufficient observation data is collected for all background
pixel values. This will significantly reduce the learning time for those games
in which foreground objects move reasonably fast, but may not help with
improving the learning efficiency for games with slow-moving objects. 

The adoption
of a large image database for each answer object could pose a challenge
to our approach since it allows for the creation of many different foreground
answer object configurations for the same game. In the worst case, a challenge
may contain none of the previously learned answer objects for that particular
game.  Continuous learning will be activated in such cases and can also be used
as a way for auto attacking in the run time. Such cases fall into the category
of ``known foreground answer objects and known target objects,'' and the success
rate can be estimated using the number of foreground objects ($o$), 
number of answer objects ($t$), and number of drag and drop attempts
allowed for each object ($a$). For example, if $o=5$, $t=3$ and $a=2$, the success rate is approximately
$\frac{2^3}{C(5,3)3!} =  13\%$.  Though as low as it seems, the rate itself is
not affected by the image database size. 

During attacking, there is
a time lapse between selecting a foreground object and verifying whether it is
an answer object. Both feature extraction and database lookup (through feature
matching) take time. In our implementation, we chose to click and hold a
selected object until a match with an answer object in the database is registered. 
In doing so, we guarantee that an answer
object, once verified, can be readily dragged and dropped, thus to avoid
dealing with the issue of constantly moving objects. However, this approach
may fail if a constraint is added by the captcha implementation that limits the amount
of time one can hold an object from moving. 
A less invasive attacking method would be to utilize parallel processing, in
which one thread is created to perform feature extraction and comparison, and
another parallel thread is used to track and predict the movement of the object
currently under verification.


\noindent {\bf \underline{Summary of Automated Attack Analysis}:} 
Our attack represents a novel approach to breaking a representative DCG captcha category.
Attacking captcha challenges, for which the knowledge
already exists in the dictionary, is 100\% accurate and has solving times 
in line with that of human users. However, building the dictionary itself is a
relatively slow process.
Although this process can be sped-up as we discussed, it may still pose a
challenge as the automated attack may need to repeatedly scan the
different captcha challenges from the server to continuously build an
up-to-date dictionary.  The defense strategies for the DCG captcha designers may
thus include: (1) incorporating a large game database as well as large object
image databases for each game; and (2) setting a lower game time-out (such as
20-30s) within which human users can finish the games but background learning
does not fully complete.  Since our attack relies on the assumption that the
background is static, another viable defense would be to incorporate a
dynamically changing background (although this may significantly hurt
usability).
It is also important to note that, as per the findings reported in 
\cite{savage-solvers}, the use of fully automated solving services represent 
economical hurdles for captcha attackers.  This applies to traditional
captchas as well as DCG captchas.
Eventually, this may make automated attacks themselves less viable in
practice \cite{savage-solvers}, and further motivates the attacker, similar to
other captchas, to switch to human-solver attacks against DCG
captchas. 


\section{Relay Attacks}
\label{sec:relay}

Human-solver relay attacks are a significant problem facing the captcha
community, and most, if not all, existing captchas are completely
vulnerable to these attacks routinely executed in the wild
\cite{savage-solvers}. In this section, we assess DCG captchas
w.r.t. such attacks. 

\subsection{Difficulty of Relaying DCG captchas}

The attacker's sole motivation behind a captcha relay attack is to completely
avoid the computational overhead and complexity involved in breaking the
captcha via automated attacks. A pure form of a relay attack, as the name
suggests, only requires the attacker to relay the captcha challenge and its
response back and forth between the server and a human-solver. For example,
relaying a textual captcha simply requires the bot to (asynchronously) send the
image containing the captcha challenge to a human-solver and forward the
corresponding response from the solver back to the server. 
%
Similarly, even video-based character recognition captchas
\cite{nu-captcha,carleton-captcha} 
can be broken via a relay attack by taking enough snapshots of the video to
cover the captcha challenge (i.e., the distorted text within the video) which
can be solved by remotely located humans. They can also be broken by simply
taking a video of the incoming frames and relaying this video to the
human-solver. 

In contrast, DCG captchas offer some
level of resistance to relay attacks, as we argue in the rest of this section.  In making this
argument, we re-emphasize that the fundamental motivating factors for a human-solver relay
attacker are simplicity and practicality. As such, a relay attack that
requires sophistication  (e.g., special software, complexity and 
overhead), is likely  not going to work in practice \cite{savage-solvers}.

There appears to be a few mechanisms using which DCG captchas could potentially
be subject to a relay attack. First, if the server sends the game code to the
client (bot) (\textit{Model 1}), the bot may simply ship the code off to the
human-solver, who can complete the game as an honest user would. However, in
\textit{Model 1}, the game code is obfuscated and can be enforced to be
executable only in a specific domain/host (e.g., only the client machine challenged
with the captcha) authorized by the server using existing
tools\footnote{http://www.kindi.com/swf-encryption.php}, which will make this
attack difficult, if not impossible. 

The second possibility, called \textit{Stream Relay}, is for the bot to employ a
streaming approach, similar to cloud gaming \cite{cloud-game}. For example, in
\textit{Model 2}, the bot
can synchronously relay the incoming game stream from the server over to the
solver, and then relay back the corresponding clicks made by the solver to the
server. Although the \textit{Stream Relay} attack might work and its
possibility can not be completely ruled out, it presents two main practical obstacles for
the attacker:

\begin{enumerate}

\item \textit{Overhead and Complexity}: The bot must be programmed (a) as a game
streaming server (which involves the overhead and complexity of processing, encoding or compressing incoming
frames), or (b) as a router which needs to ``sniff'' raw networking packets (which may not
be permitted on a compromised non-root user account).  ((a) and (b) may also be
detected by an IDS running on the machine compromised by bot)

\item \textit{Latencies}: Streaming or routing a large number of game frames
over a (usually) slow connection between the bot (e.g., based in the US) and
the solver's machine (e.g., based in China) would degrade the game performance
(similar to video streaming over slow connections), reducing solving accuracy
and increasing response time. Such differences from an average honest user game play session
may further be used to detect the attack. 

\end{enumerate}	

It is also important to note that the  Stream Relay attack represents a significant shift
from a traditional captcha relay attack, which may defeat the motivation
for pursuing a relay attack in the first place.

This motivates us to consider another much simpler relay attack approach called \textit{Static
Relay}. Here, the bot asynchronously relays a \textit{static snapshot} of the game to a
human-solver and uses the responses (locations of answer objects and that of the
target objects) from the solver to break the captcha (i.e., drag and drop the
object locations provided by the solver to the target object locations provided
by the solver).

\begin{figure}[h]
\begin{center}
\vspace{-3mm}
\includegraphics[width=\columnwidth]{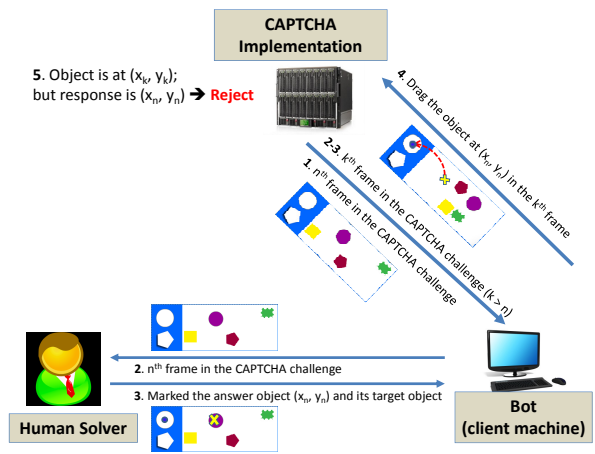} 
\vspace{-5mm}
\caption{Difficulty of succeeding at the Static Relay attack}\label{fig:relay} 
\vspace{-2mm}
\end{center}
\end{figure}

The Static Relay attack approach is very simple and in line with a traditional
captcha attack (and thus represents a viable relay attack). However, it is expected
to have poor success rates.  The intuitive reason behind this is a natural
\textit{loss of synchronization} between the bot and the solver, due to the
dynamic nature of DCG captchas (moving objects).  In other words, by the time
the solver provides the locations of target object and the answer objects
within a challenge image (let us call this the $n^{th}$ frame), the objects
themselves would have moved in the subsequent, $k^{th}$, frame ($k>n$), making
the prior responses from the solver of no use for the bot corresponding to the
$k^{th}$ frame.  Recall that the objects move in random directions and often
collide with other objects and game border, and therefore it would not be
possible for the bot to predict the location of an object in the $k^{th}$ frame
given the locations of that object in the $n^{th}$ frame ($n<k$).  Such a loss
of synchronization will occur due to: (1) communication delay between the bot
and human solver's machine, and (2) the manual delay introduced by the solver
him/herself in responding to the received challenge.  Figure \ref{fig:relay}
shows such a desynchronization in attempting to relay a DCG captcha.

A determined Static Relay attacker (bot) against the DCG captcha can, however, attempt to
maximize the level of synchronization with the solver.
Although it may not
possible for the attacker to minimize (ideally, eliminate) the communication
delays
(especially for bots potentially thousand of miles away from human solvers),
it may be possible to minimize the manual delay via the introduction of
carefully crafted tasks for the human-solver. In the rest of this section, we
report on an experiment and the results of an underlying user study in order to
evaluate the feasibility of Static Relay attack against our DCG captcha instances.
This novel experiment takes the form of a \textit{reaction time} or \textit{reflex
action} task for the human-solver. A reaction time task involves performing
some operation as soon as a stimulus is provided. A common example is an
athlete starting a race as quickly as a pistol is shot.
The subject of reaction time has been extensively studied by psychologists (see
Kosinski's survey
\cite{kosinski}).

\subsection{Reaction Time Static Relay Experiment}
\label{sec:relay-exp}


Our hypothesis is that DCG captchas will be resistant to the \textit{Static Relay} attack, and so we
  give the attacker a strong power in the following sense: our tests
  eliminate the communication delay between the bot and the human solver,
  by putting them on the same machine. The focus of the experiment
  then
%
shifts towards  motivating
human-solvers to perform at their best by employing meaningful interfaces and
by framing the underlying task in a way that is amenable to these solvers.  In
particular, since attacker's goal is to minimize the delay incurred
by the human solver in responding to the challenges,  we model human-solver attack as
a reaction time \cite{kosinski} task described below. 
Our Section  \ref{sec:relay-user} study further facilitates the attacker with human
  solvers having low response times and quick reflex actions, such as youths
  in their 20s \cite{kosinski}. 
 


\vspace{1mm}
\noindent\textbf{Experimental Steps:} The reaction time Static Relay attack experiment
consists of the following steps:

\begin{enumerate}

%
\item A snapshot of the game challenge is extracted by the bot (B), and the
human solver (H) is asked to identify/mark a target object for that
game challenge (e.g., the dog in the Shapes game). 

%
\item For each target object identified above, H is asked to identify one
answer object in the snapshot specific to the game challenge (e.g., bone for
the dog in the Shapes game). However, since B wants to minimize the delay
between the time the challenge snapshot is given and the response is received
from H, a stimulus will be associated with the snapshot. We make use of a
combination of (1) a visual stimulus (the border across the game window flashes
in Red) and (2) an audio stimulus (a beeping sound).  The task for H is to
identify an answer object in the image \textit{as soon as} the stimuli are provided. 

%
\item B will emulate the dragging and dropping of the objects based on the response of H (simply use
the pixel values provided by H as the coordinates of the objects and respective targets). 

%
\item Steps 2-3 are repeated until all answer objects for a given target object are identified by H and
dragged/dropped by B.

\item Step 1 is repeated until all target objects have been covered.

\end{enumerate}

The experiment succeeds if the captcha game completes successfully, i.e.,
if all answer objects are dragged to their respective targets by B per input from H.  



\vspace{1mm}
\noindent\textbf{Experimental Implementation:}	
%
Our implementation of the above experiment consists of a user interface (UI) developed in Java that
interacts with the human solver and a bot. The core of this implementation is
designed using an algorithm following which the screen
captures are updated and displayed on the screen as well as an algorithm used
to make the mouse drag and drop of the objects. 

The game starts by the bot capturing an image of the game challenge from the
browser (i.e., the captcha challenge that the bot received from the server) and
displays that image in the UI. The solver is then asked to click on a target
object within that image. After selecting the target, the solver is instructed
to click a ``Next'' button, wait for a flashing and a beep (our stimuli),
followed by clicking the object that matches with that target. Once the solver
has clicked on the object, the bot takes control of the mouse by clicking and
dragging the object to the target in the flash game. The solver must be able to
identify and choose the correct object before the object has moved too far in
the flash game displayed in the browser.  Whether the click is successful or
not, a new screen capture is retrieved from the game on the browser. If the
solver has chosen the object in time on the UI, then he/she can pick a new
target if one exists by clicking on the ``New Target'' button. If the solver
has missed clicking on the object fast enough (i.e., if the click was not
successful), the solver will automatically get another attempt to choose the
correct object followed by the flashing and the beep. Figure
\ref{fig:relay-ui}, Appendix \ref{sec:app-fig} depicts the UI of our implementation.


\subsection{Static Relay Attack User Study}
\label{sec:relay-user}

%

We now report on a user study of the aforementioned reaction time relay attack
experiment presented in Section \ref{sec:relay-exp}.  

\subsubsection{Study Design, Goals and Process}

In the relay attack study, users were given the task to play our 4 game instances through
the UI (described above). 
The study comprised of \textbf{20 participants}, primarily Computer Science 
university students. This sample represents a near ideal
scenario for an attacker, given that young people typically have fast reaction
times \cite{kosinski}, presumably optimizing the likelihood of the success of the
relay attack. The demographics of the participants are shown in
Table~\ref{tab:relay-demographics-interviews}. The study design was
similar to the one used in our usability study (Section \ref{sec:usability}).
It comprised of three phases.  The \textit{pre-study phase} involved
registering and briefly explaining the participants about the protocols of the
study. 
This was followed by
collecting participant demographics and then the participants playing the games
via our interface.  
The participants were told to perform at their best in playing the games.
The \textit{post-study phase} required the participants to respond to a set of 
questions related to their experience with the games they played as part
of the interface, including the SUS questions \cite{sus}. 


\begin{table}[htdp]
{\footnotesize{
\caption{Relay Attck User Study Participant Demographics}
\begin{center}
\small
\begin{tabular}{|p{5cm}| r |}
\hline
\textbf{}& \textbf{N=20} \\
\hline
\hline
\textbf{Gender} & \%\\
\hline
 Male &70 \\
 \hline
Female&30 \\
\hline
 \hline
\textbf{Age} & \% \\
\hline
18 - 24 &35 \\
 \hline
 25 - 35 &60 \\
  \hline
35 - 50  &5\\
  \hline
  \hline
\textbf{Education} & \% \\
\hline 
Highschool &25 \\
 \hline
Bachelors  &45 \\
 \hline
Masters &30 \\
\hline
Ph. D. & 0 \\
 \hline
  \hline
\textbf{Profession / field of study} & \% \\
  \hline
Computer Science & 90\\
  \hline
Engineering &5 \\
 \hline
Medicine & 5 \\
 \hline
 \end{tabular}
\normalsize
\end{center}
\label{tab:relay-demographics-interviews}
}}
\end{table}

Each participant was
  asked to play the relay versions corresponding to each of the 20
  variations of the 4 DCG captcha games as in Section \ref{sec:design}; we used
  ordering based on Latin squares, as in the usability study.
The specific goal of our study was to evaluate the reaction time experiment UI in terms of the following aspects:


\begin{enumerate}
\item {\textit{Efficiency}: time taken to complete the games (and succeeding at the relay attack)}.
\item {\textit{Robustness}: likelihood of not completing the game (relay attack failure), and incorrect drags/drops.}
\item {\textit{User Experience}: quantitative SUS ratings and qualitative feedback from the participants.}
\item  {\textit{Reaction time}: Time delay between the presentation of the stimuli and the response from the participant. This is a fundamental metric for the feasibility of the attack. If reaction time is large, the likelihood of attack success will be low.}
\end{enumerate}

Another important goal of our user study was to compare its performance with
that of the usability study. If the two differ significantly, the 
relay attack can be detected based on this difference.

For each tested game, completion times and errors
were automatically logged by the our web-interface software. In
addition, we maintained ``local logs'' of the clicks made by the participants
on our game interface to measure the reaction timings.

\subsubsection{Study Results}

We present various mechanical data, i.e., time to completion and error rates
as part of the relay attack study. We further analyze the local logs for the reaction
time analysis.
%


\noindent {\bf Completion Time and Error Rates}: Table
\ref{tab:relcaptcha} shows the time taken and error rates to play the games for
each game type by different participants. Unlike our usability study, many
game instances timed out, i.e., the participants were not able to always complete
these game instances within the time out of 60s.  In this light, we
report two types of timings: (1) \textit{successful time}, which is the time only
corresponding to the games that the participants were able to complete
successfully within the time out, and (2) \textit{overall time}, which is the
time corresponding to both the game instances completed successfully within the
time out and those which timed out (in which case we consider the timing to be
60s). The overall time therefore will effectively be higher. 

All games turned out to be quite slow, and much slower than that of
the usability study where the games lasted for less than 10s on an
average (Section \ref{sec:usability}).  
As in our usability study, we found that users took longest to solve the Animals
  (overall average: 46.51s), whereas
the other games took slightly less time.
This might have been due to the increased semantic load in the Animals game due to the
presence of 3 target objects.  We observed that the error
rates were the highest for the Animals game (40\%), and the least for the
Shapes games (9\%) although the corresponding per click error rates were
high (56\%).  The Ships and Parking games had comparable overall error rates
between 20-30\%.
We analyzed and further compared the mean time for different game categories. Using the ANOVA test, the
games showed statistically different behavior from each other ($F=12.85$,
$p<0.0001$). On further analyzing the data, we found the following pairs of
games to 
be statistically different from each other:
Shapes and Ships ($p=0.027$)
and Animals and all other games
($p<0.001$). 

To analyze errors better, we investigated error rates per click, i.e.
for  each drag attempt whether the object being dragged was dropped at the
correct position or not. The error rate per click was the least for the Ships
game (17\%), much lower compared to all other games (50-70\%), 
the latter itself being much higher than
  observed during the usability study. This suggests that the server
  could prevent 
 the
relay attack against Animals, Parking and Shapes games by simply capping the number of drag/drop attempts.


{\footnotesize{
\begin{table}[ht]
  \begin{center} 
  \caption{Error rates and completion time per game type}
    \begin{tabular}{|p{.9cm}|p{1.9cm}|p{1.9cm}|p{.7cm}|p{1.6cm}|}

   \hline
{\bf Game Type} &  \bf Overall Time (s)& \bf Successful time (s) &\bf Error Rate &\bf Error Rate Per click   \\
&  \textit{mean (std dev)}& \textit{mean (std dev)}&\textit{mean}& \textit{mean}   \\
 \hline
 \hline
    \raggedright{Ships} &30.92 (5.91) & 22.25 (5.04) & 0.26  &0.17 \\
 \hline
  \raggedright{Animals} &46.51 (5.05)&37.93 (4.91) & 0.40  &0.65  \\
\hline
  \raggedright{Parking} &28.16 (7.36) &20.45 (5.04) & 0.22  & 0.66  \\
  \hline
  \raggedright{Shapes} &26.19 (1.59)  &22.94 (1.74) &0.09 &  0.56 \\
\hline 
    \end{tabular}
  \label{tab:relcaptcha}
\end{center} 
\end{table}%
}}





\noindent {\bf Reaction Time}:
%
We now analyze the reaction time corresponding to different games during the
relay attack experiments.  We consider two types of reaction times, one
corresponding to all clicks made by the participants, and the other
corresponding to only the correct clicks (i.e., those that resulted in a
correct drag and drop).  The averaged results for the two types of reaction
times for each game type are summarized in Table \ref{tab:relay1}.  We can see
that the average reaction time (all clicks) for all game categories was more
than 2s and the least for the Shapes game (2.17s). The average
reaction time (correct clicks) is slightly lower than reaction time (all
clicks), but still higher than 1.5s and still lowest for the Shapes game (1.62s). 
Neither types of reaction
times change significantly across different game categories. 
ANOVA test, however, did find
significant difference between the mean reaction time (all clicks) of the four games ($F=13.19$, $p<0.01$). On further analyses using paired t-tests with Bonferroni correction,
we found that there was a significant difference between the Animals
and Parking games ($p<0.01$). 
Similarly, using the ANOVA test, we found
significant difference between the mean reaction time (correct clicks) ($F=3.24$, $p<0.027$). Further,
we found a significant difference between the Shape
and Ship games ($p<0.005$) with respect to mean reaction time (correct clicks).


%


{\small{
\begin{table}[ht]
  \begin{center}
  \caption{Reaction times per game type}
    \begin{tabular}{|p{1.25cm}|p{2cm}|p{2.4cm}|}
   \hline
{\bf Game Type} &  \bf Reaction Time All Clicks (s) & \bf Reaction Time Correct Clicks (s)\\
&  \textit{mean (std dev)}& \textit{mean (std dev)}  \\
 \hline
 \hline
    \raggedright{Ships} &2.27 (0.34)  &2.06 (0.17)\\
 \hline
  \raggedright{Animals} & 2.58 (0.35) & 1.85 (0.23) \\
\hline
  \raggedright{Parking} & 2.50 (0.51) & 2.00 (0.31)\\
  \hline
  \raggedright{Shapes} &2.17 (0.2) &1.62 (0.11)\\
\hline 
    \end{tabular}
  \label{tab:relay1}%
  \end{center}
\end{table}
}}

\noindent {\bf User Experience}:
%
We now consider the data collected via direct user responses during the
post-study phase.  The average SUS score from the study came out to be only
$49.88$ (std dev = $5.29$). This is rather low given that average scores for
commerical usable systems range from 60-70 \cite{susmean}, and suggests a poor
usability of the system. This means that it would be difficult for human users
to perform well at the relay attack task and implies that launching relay
attacks against DCG captchas can be quite challenging for an attacker.


Table~\ref{tab:relaysus} shows the 5-point Likert scores (`1' is ``Strong Disagreement''; `5' is ``Strong Agreement'') for the visual appeal and
pleasurability of the games. Although the former average ratings are
on the positive side (more that 3), the latter ratings are low, suggesting the participants
did not find the games to be pleasurable.  In our games, we made use of
visual and audio stimuli to which the users had to respond.  In order to
understand what type of stimulus worked best for the participants, we asked
them to what extent the audio, visual or both stimuli together was useful as an
indicator to respond fastest to the game.  These ratings are depicted in
Table~\ref{tab:relaysus}.  The responses were on average in favor of the visual
stimulus, followed by the two stimuli together, and finally the audio stimulus.
35\% participants found audio stimulus and visual stimulus to be sufficient
whereas 45\% participants agreed or strongly agreed with the statement that
both visual and audio stimulus are necessary to play the game.  We further
performed the ANOVA test for responses corresponding to the three --
visual, audio and visual+audio -- stimuli, but did not find a statistically significant difference. Finally, 80\% of the
participants felt that training will help them play the games better with an
average score of 3.95. 
This suggests an attacker might
  improve success in relay attack through advance training of human
  solvers.

{\small{
\begin{table}[ht]
  \begin{center}
  \caption{Participant Feedback Summary}
    \begin{tabular}{|p{3cm}|p{3cm}|}
   \hline
{\bf Features} &  \bf Likert Mean (std dev)\\
 \hline
 \hline
    \raggedright{Visually Attractive} & 3.20 (0.92)\\
 \hline
  \raggedright{Pleasurable } & 2.85 (0.99)\\
\hline
  \raggedright{Visual Stimulus } &3.20 (1.17)\\
  \hline
   \raggedright{Audio Stimulus } &2.95 (0.93)\\
  \hline
    \raggedright{Both Audio and Visual } &3.10 (1.25)\\
  \hline
    \raggedright{Need Training } &3.95 (1.01)\\
  \hline
    \end{tabular}
  \label{tab:relaysus}%
  \end{center}
\end{table}%
}}


\vspace{1mm}
\noindent {\bf \underline{Summary of Relay Attack Analysis}}: Our analysis
suggests that subjecting the DCG captcha to relay attacks poses significant
challenges. Either the relay attack becomes complex for the bot (e.g., \textit{Stream
Relay}) or it remains to be very simple for the bot but becomes very difficult for the human-solver (\textit{Static Relay}).  
Specifically, for the {Static Relay} attack to succeed, it is necessary for the human solver to
perform a reaction time task (average reaction time is more than 2s).  This
task, except for the Shapes game, takes much longer ($>$ about 30s on average),
is significantly more error prone (error rates more than 20\%; per click error
rates more than 50\%), and much harder for the users \textit{when compared} to
directly playing the games by honest users under a non-relay attack setting.
These numbers only represent a \textit{lower bound}.  In real life, where the
communication delays between the bot and solver's machine will be non-zero, and
average solver population samples are used (unlike our attack set-up), the
timings and error rates will be higher and launching a relay attack would be
even more difficult. Although our experiments were conducted on our 4 DCG
captcha instances, we believe that our analysis is generally applicable to
other DCG captcha types involving moving answer objects.



\section{Related Work}
\label{sec:related}

\begin{figure}[h]
\begin{center}
\vspace{-6mm}
{\includegraphics[width=\columnwidth]{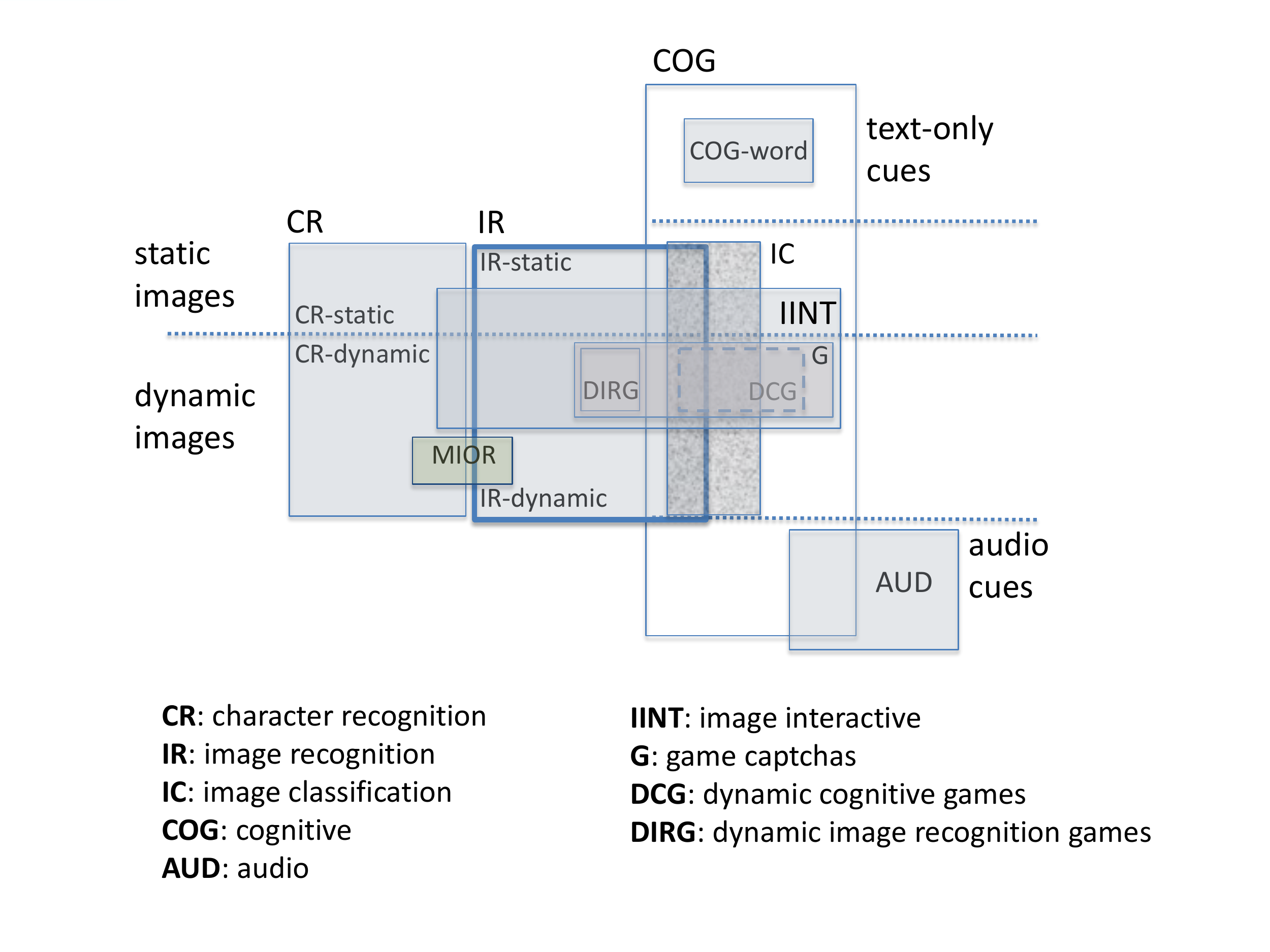}} 
\vspace{-10mm}
\caption{Captcha Taxonomy} \label{fig:tax}
\vspace{-2mm}
\end{center}
\end{figure}

%
In this section, we review prior work and provide a taxonomy on captchas. For details, we refer the reader to 
~\cite{carleton-captcha}, 
~\cite{basso10}, 
~\cite{hidalgo:captchas}. Figure \ref{fig:tax} depicts the \capt taxonomy extending earlier classifications

Character-recognition (CR)
\capts are the most common today,  requiring users to input alphanumeric characters corresponding to
distorted or obscured strings.  Attacks~\cite{Bursztein:2011,YanA07,Yan:2011}
typically
involve optical character recognition tools, with security relying heavily on
the difficulty of character segmentation.
Image recognition (IR) captchas \cite{Zhu:2010}
involve images or objects, other than alphanumeric characters, 
with the user requiring to identify or
recognize certain objects as distinct from others.
For example in Asirra \cite{Elson:2007, Golle:2008},
cats are distinguished from dogs.
We identify 
image classification (IC) \capts as a separate category -- in many ways similar to IR \capts,
but with main task being \textit{classification}
(involving cognitive effort or reasoning based on
image semantics) more than simple \textit{recognition}.
This is motivated by experience in
computer vision indicating 
classification to be much more difficult than recognition \cite{carleton-captcha}.

Captchas involving images (including of text) can be sub-divided based on
the images being static or dynamic.
The moving-image object recognition (MIOR) 
sub-category~\cite{carleton-captcha} 
includes, e.g., NuCaptcha~\cite{nu-captcha}
and emergent images~\cite{emerging:2009}.   
IC \capts are a sub-category
of cognitive (COG) \capts, 
which include
text-based puzzles (e.g., leveraging language constructs or word semantics), 
logic questions, and semantic tasks related to images (e.g., content-based tagging of
YouTube videos \cite{video_captcha}.   
Naor~\cite{Naor96} anticipated many COG schemes. 

Many recent proposals are image-interactive (IINT) captchas, wherein users
interact with static or dynamic images by clicks, screen touches, or
drag/drop actions (e.g., \cite{zhang-arch,dd1,dd2,sketcha}).  Of particular
interest to our present work is the game-based (G) \capt subcategory, in which
we identify further sub-classes: dynamic cognitive games (DCG), and dynamic
image-recognition games (DIRG).  Motivated by vision-impaired users, audio
\capts (AUD) \cite{Bursztein:2011b,Bursztein:2009,Soupionis:2010}, is another
major category, but it suffers both usability and security problems
\cite{Bursztein:2011b,Bursztein:2009}.

The \capt literature includes work on robustness~\cite{chellapilla,Chellapilla:2005hipworkshop,Zhu:2010}  
and guidelines ~\cite{Bursztein:2011,hidalgo:captchas,captcha_usability}.   
Studies on the outsourcing of \capt solutions, cheap labor,
\capt farms, and the use of Mechanical Turk~\cite{bursztein:how-good-are-humans-at-so:2010:zlygf,egle10,savage-solvers}
lead to serious questions about limitations---\capts were originally
intended to distinguish humans from computers, not 
legitimate users from other humans willing to cheaply solve them, or tricked into solving
them through relay attacks. 

\section{Conclusions and Future Work} \label{sec:conclude}
This paper represents the first academic effort towards investigating the
security and usability of game-oriented captchas in general and DCG captchas in
particular --- a recent approach aimed at eliminating user frustration common
to traditional captchas.
Our overall findings are mixed.  On the positive side, our results suggest that
DCG captchas, unlike other known captchas, offer some level of resistance to
relay attacks. 
We believe this to be a significant advantage of these captchas given that
other captchas are routinely broken via relaying.
Furthermore, the studied representative DCG captcha category demonstrated high
usability. On the negative side, however, we have also shown this category to
be vulnerable to a dictionary-based automated attack.  
Prior to our work, the community view \cite{savage-solvers} for captchas in
general was that fully automated attacks are --- due to economic reasons ---
less appealing to real-world captcha attackers than human-solver relay attacks.
In contrast, our work shows that for DCG captchas, the opposite might be true,
due to the difficulty of relaying DCG captchas --- thus making these captchas
an interesting area for further study and evolution. 

An immediate consequence of the insights borne out from our study is that
further research on DCG captchas could concentrate on making these captchas
better resistant to automated attacks while maintaining a good level of
usability.\footnote{Any modifications
made to the original DCG captchas to resist automated attacks, such as a
dynamic background, may further make relay attacks more difficult.} Moreover,
our paper focused on ``pure automated'' and ``pure relay attacks.'' However,
different forms of hybrid attacks can be envisioned, which carefully combine the
computing power and human knowledge to undermine the security of DCG captchas.
Future research is necessary to investigate how well hybrid attacks work, and
how they alter the economics of captcha-solving (following up
\cite{savage-solvers}). Finally, the results of our user study on reaction-time
task performance may have general applications in human-centered computing
(security and non-security) domains.  For instance, these results may rule out the
possibility of usable captcha schemes themselves based on a reaction-time test.

\section*{Acknowledgments}

We thank: the team of ``are you a human'' for creating the CAPTCHAs that
inspired our work; John Grimes, John Sloan and Anthony Skjellum for guiding us
regarding the ethical aspects of our work and advising us on pursuing our own
implementation; Sonia Chiasson, Fabian Monrose and Gerardo Reynaga regarding
early discussions; all members of the SPIES group at UAB and
PreCog group at IIITD for helpful suggestions throughout the study; and all
participants of our usability and user studies. The work of Mohamed, Georgescu
and Saxena is directly supported by a grant on ``Playful Security'' from the
National Science Foundation CNS-1255919. Van Oorschot holds the Canada Research
Chair in Authentication and Computer Security and acknowledges NSERC for
funding the chair and a Discovery Grant.

{\footnotesize{

}}

\appendix
\section*{A. Additional Figures and Tables}
\label{sec:app-fig}

\begin{figure}[htbp]
\begin{center}
\centering
\subfigure[A star indicating correct object match]{
	\includegraphics[width=.21\textwidth]{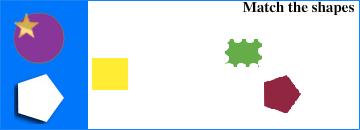}
  	\label{fig:star}
 }
\centering
\subfigure[A cross indicating incorrect object match]{
	\includegraphics[width=.21\textwidth]{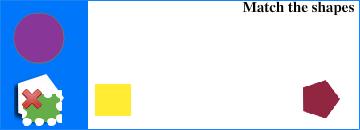}
  	\label{fig:cross}
 }
\vspace{-4mm}
\caption{User Feedback per Game Interaction}
\vspace{-5mm}
\label{fig:feedback}
\end{center}
\end{figure}


\begin{figure}[h]
\vspace{-.2in}
\begin{center}
\centering
\subfigure[The solver is asked to choose a target object]{
	\includegraphics[width=.25\textwidth]{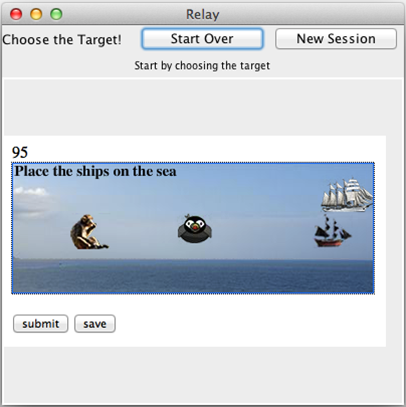}
  	\label{fig:start}
 }
\hspace{.5in}
\centering
\subfigure[The solver is asked to choose the next answer object, if any]{
	\includegraphics[width=.25\textwidth]{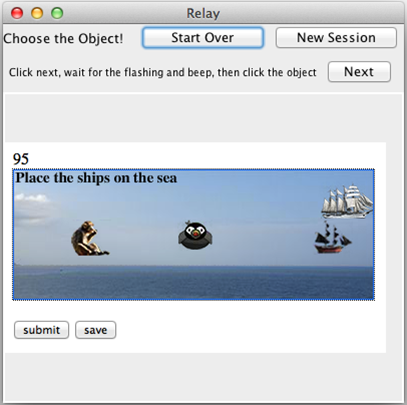}
  	\label{fig:next}
 }
\centering
\hspace{1in}
\subfigure[The solver is asked to select a new target object, if any]{
	\includegraphics[width=.25\textwidth]{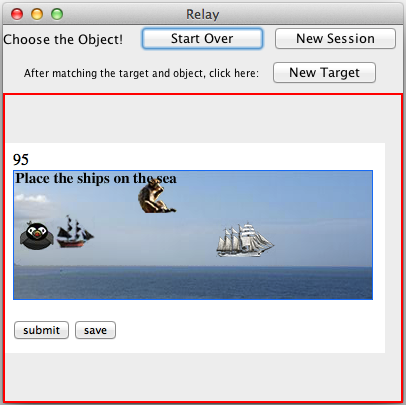}
  	\label{fig:new}
 }
\vspace{-3mm}
\caption{User interface implementing the reaction time relay experiment (95 represents the User ID; the red rectangle in (c) represents our visual stimulus)}
\vspace{-6mm}
\label{fig:relay-ui}
\end{center}
\end{figure}

\section*{B. Target Area Detection using the Exclusion Method}
\label{sec:exclusion}

\begin{figure}[h]
\begin{center}
\vspace{-.5in}
\includegraphics[width=\columnwidth]{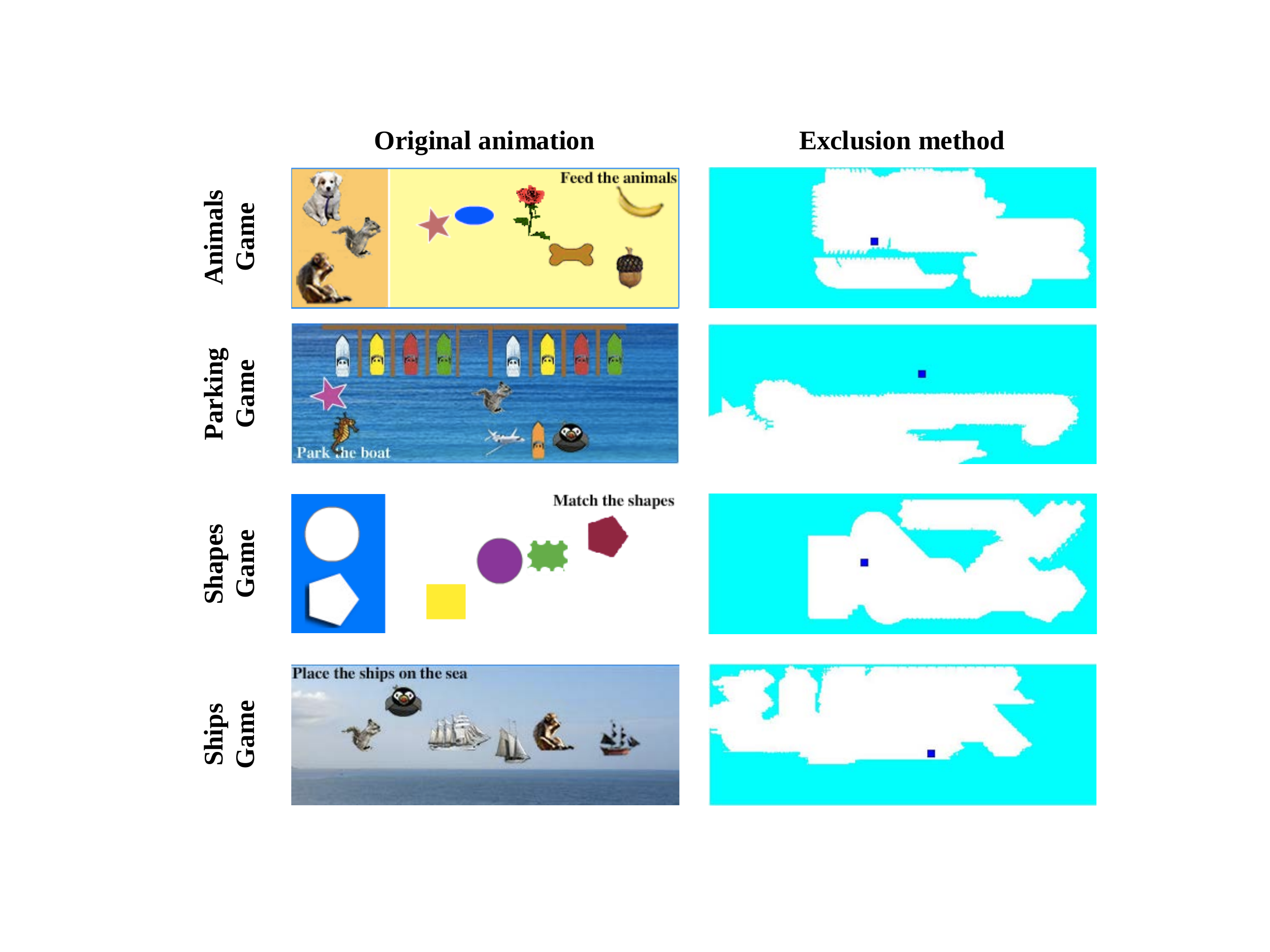} 
\vspace{-.6in}
\caption{The target area centers detected by exclusion method }
\vspace{-1mm}
\label{fig:target-center}
\end{center}
\end{figure}

A design alternative for target area detection, called the \textit{exclusion method} 
is to detect the target area by simply removing foreground object pixels
accumulated from all the sample frames. However, while this alternative is
slightly faster but still at about the same time efficiency as the MBR-based
method, it is less robust than the latter especially when the objects are
moving slow such that the remaining area, i.e., the detected target area, may
include too much of the foreground object moving area that has not had a chance
to be covered by the footprints of foreground objects extracted from the
limited set of sample frames. Figure \ref{fig:target-center} shows our experimental results for this
design alternative applied to four different challenges, where the blue dots
represent the detected target area centers. This alternative method failed to
detect the correct target center in all four cases.

\end{document}